\documentclass[3p,times,procedia]{elsarticle}
\usepackage{nupha_ecrc}

\volume{00}

\firstpage{1}

\journalname{Nuclear Physics A}

\runauth{}

\jid{nupha}

\jnltitlelogo{Nuclear Physics A}

\usepackage{amssymb}
\usepackage{amsmath}

\usepackage[usenames]{color}
\usepackage[colorlinks=false,linktocpage=true]{hyperref}

\usepackage{color}
\definecolor{blue}{RGB}{0,50,230}

\usepackage{hyperref}
\hypersetup{
  colorlinks=true,
  linkcolor=blue,
  citecolor=blue,
  urlcolor=blue
}

\usepackage[figuresright]{rotating}

\newcommand{\ed}{{\cal E}}
\newcommand{\piT}{\tilde{\pi}_\perp}
\newcommand{\pit}{\tilde{\pi}}
\newcommand{\pL}{\hat{\cal P}_L}
\newcommand{\pT}{{\cal P}_{\perp}}
\newcommand{\Pit}{\tilde{\Pi}}
\newcommand{\peq}{{\cal P}_0}
\newcommand{\IT}{\tilde{I}}
\newcommand{\WT}{\tilde{W}}
\newcommand{\sigmat}{\tilde{\sigma}}
\newcommand{\thetat}{\tilde{\theta}}

\begin{document}

\begin{frontmatter}



\dochead{XXVIth International Conference on Ultrarelativistic Nucleus-Nucleus Collisions\\ (Quark Matter 2017)}

\title{Optimized fluid dynamics for heavy ion collisions$^1$}\footnote{%
Supported by the US Department of Energy under awards \rm{DE-SC0004286} and \rm{DE-SC0013470} and by the National Science Foundation through the JETSCAPE Collaboration (award number ACI-1550223).}

\author[OSU]{Dennis Bazow}
\author[OSU]{Ulrich Heinz}
\author[Kent]{Michael Strickland}

\address[OSU]{Department of Physics, The Ohio State University, Columbus, Ohio 43210-1117, USA}
\address[Kent]{Department of Physics, Kent State University, Kent, Ohio 44242, USA}

\begin{abstract}
The (viscous) anisotropic hydrodynamic approach, especially after perturbative inclusion of all residual viscous terms, has been shown to dramatically outperform viscous hydrodynamics in several simplified situations for which exact solutions exist but which share with realistic expansion scenarios the problem of large dissipative currents. We will report on the present status of applying viscous anisotropic hydrodynamics in a highly efficient simulation of the full three-dimensional quark-gluon plasma. Results from accelerated $3{+}1$-dimensional viscous hydrodynamic simulations using graphics processing units will be compared to the anisotropic frameworks.
\end{abstract}


\end{frontmatter}


\section{Introduction}
\label{intro}

Quantitative modeling of relativistic heavy ion collisions has dramatically improved in recent years. Simulations include an initial state and pre-equilibrium model for the first ${\sim\,}0.5-1$\,fm/$c$ when the quark-gluon plasma (QGP) is highly momentum anisotropic and far from thermalized, followed by viscous hydrodynamic simulations of the medium for the semi-isotropic and semi-thermal QGP for the next ${\cal O}(10)$\,fm/$c$, and finally a microscopic hadronic treatment of the reformed hadrons. One of the main uncertainties in this chain is from the early evolution stage when the QGP is highly anisotropic and cannot be treated with viscous hydrodynamics. This uncertainty is usually encapsulated in model parameters describing the initial state and early pre-equilibrium dynamics. To constrain these and other model parameters with the help of experimental observables, these simulations are coupled with advanced statistical methods based on Bayesian statistics \cite{Novak:2013bqa,Bernhard:2016tnd}. Since it is very computationally expensive to train these models to cover very high-dimensional parameter spaces we optimize the simulations along two fronts: we (1) speed up the $3{+}1$ dimensional fluid dynamic simulation by performing the calculations on graphics processing units (GPUs) \cite{Bazow:2016yra} and (2) limit the uncertainty from the early evolution stage by using the viscous anisotropic hydrodynamic formalism \cite{Bazow:2013ifa} which accounts for the large momentum anisotropies at early times non-perturbatively and thus allows us to start the hydrodynamic stage earlier. The viscous anisotropic formalism improves upon leading-order anisotropic hydrodynamics (see \cite{Strickland:2014pga} for a review) by also including previously neglected residual components of the shear stress tensor through Isreal-Stewart-like perturbative transport equations. The combination of these advances leads to faster and quantitatively more reliable dynamical simulations of heavy ion collisions with fewer parameters.

\section{GPU-accelerated (3+1)-dimensional second-order viscous hydrodynamics}
\label{perf}

Our implementation of second-order viscous relativistic fluid dynamics on graphics processing units (GPU-VH) is described in \cite{Bazow:2016yra}. This code has since been adapted to evolve the second-order anisotropic hydrodynamic equations of motion, but not yet on GPUs. The performance of GPU-VH, at different spatial resolutions, is measured via the time it takes to complete 100 full time steps. Table~\ref{T1} compares performance of GPU-VH on the GeForce GTX 980 Ti graphics card relative to the CPU-VH code run on the host machine with a 2.6 GHz Intel Xeon CPU E5-2697 v3 using a single core. We observe speed-up factors of ${\cal O}(100)$. 
\begin{table}[t!]
\small
\centering
\begin{minipage}{0.5\linewidth}
\begin{tabular}{|c|c|c|c|}
  \hline
  Number of grid points & C/CPU & CUDA/GPU & Speedup \\
  & (ms/step) & (ms/step) & \\
  \hline
  128 $\times$ 128 $\times$ 32 	& 7145.978  & 63.261 & 112.960 \\
  128 $\times$ 128 $\times$ 64 	& 13937.896 & 123.527 & 112.833 \\
  128 $\times$ 128 $\times$ 128 	& 30717.367 & 244.450 & 125.659 \\
  256 $\times$ 256 $\times$ 32 	& 25934.547 & 236.593 & 109.617 \\
  256 $\times$ 256 $\times$ 64 	& 57387.141 & 472.391 & 121.482  \\
  256 $\times$ 256 $\times$ 128 	& 129239.959 & 939.340 & 137.586 \\   
  \hline
\end{tabular}
\end{minipage}
\hspace*{0.13\linewidth}
\begin{minipage}{0.25\linewidth}
\caption{Performance results of the C/CPU and CUDA/GPU versions of CPU-VH and GPU-VH by measuring the computer time it takes to complete one full RK step, averaged over 100 time steps, at different spatial resolutions.  v3.}
\label{T1}
\end{minipage}
\vspace*{-3mm}
\end{table}

\section{(3+1)-dimensional second-order anisotropic hydrodynamics}
\label{vahydro}

Assuming negligible net baryon density, relativistic fluid dynamics is described by the conservation laws for energy and momentum, $\partial_{\mu}T^{\mu\nu}(x){\,=\,}0$, complemented by relaxation-type evolution equations for the dissipative flows. For anisotropic systems $T^{\mu\nu}$ can be decomposed with respect to the fluid four-velocity $u^\mu$ and the space-like four-vector $z^\mu$ (defining the direction of the largest anisotropy, which for heavy-ion collisions is the beam direction \cite{Strickland:2014pga}), parametrized by $u^\mu\equiv(u^0,u^1,u^2,u^3) = (u_\tau,\vec{u}_\perp,u_\eta)$ and $z^\mu=\gamma_{z}(\tau u^3,0,0,u_\tau/\tau)$, where $\gamma_z^{-2}\equiv 1+u_\perp^2$. Identifying the energy density $\ed$ with its equilibrium form via Landau matching and demanding that the longitudinal pressure in the direction of the anisotropy is equal to its ``anisotropic equilibrium" value \cite{Molnar:2016gwq}, we can decompose the energy-momentum tensor as (indicating ``anisotropic equilibrium" quantities with an over-hat and $\tilde{\cal O}={\cal O}-\hat{\cal O}$)~\cite{Molnar:2016vvu}:
\begin{equation}
T^{\mu\nu}=\ed u^{\mu}u^{\nu}-{\cal P}_\perp\Delta_\perp^{\mu\nu}+\pL z^{\mu}z^{\nu}+2\tilde{W}^{(\mu}_{\perp z}z^{\nu)}+\pit_\perp^{\mu\nu}\;.
\label{Tmunu_vah_uz}
\end{equation}
Here, $\pL$ is the total longitudinal pressure, the transverse pressure $\pT\equiv\hat{\cal P}_\perp+3\Pit/2$ is the sum of the ``anisotropic equilibrium" pressure $\hat{\cal P}_\perp$ and the residual bulk viscous pressure $\Pit$, $\tilde{W}^{\mu}_{\perp z}\equiv-\Delta^{\mu}_{\perp,\alpha}T^{\alpha\beta}z_{\beta}$ is the energy-momentum diffusion current in the $z$ direction, and the transverse shear stress tensor is $\pit_\perp^{\mu\nu}\equiv T^{\lbrace\mu\nu\rbrace}$. The transverse projection tensor $\Delta_\perp^{\mu\nu}\equiv g^{\mu\nu}-u^{\mu}u^{\nu}+z^{\mu}z^{\nu}$ is used to project four-vectors and tensors into the space orthogonal to $u^\mu$ and $z^\mu$. By construction, the dissipative terms satisfy the constraints $u_{\mu}\tilde{W}^{\mu}_{\perp z}=z_{\mu}\tilde{W}^{\mu}_{\perp z}=u_{\mu}\pit_\perp^{\mu\nu}=z_{\mu}\pit_\perp^{\mu\nu}=g_{\mu\nu}\pi^{\mu\nu}\equiv 0$. 

To close the conservation laws, additional evolution equations for $\pL$ and $\Pit$ (entering in Eq.~(\ref{Tmunu_vah_uz}) through ${\cal P}_\perp$), and for the residual dissipative currents $\tilde{W}^{\mu}_{\perp z}$ and $\pit^{\mu\nu}_{\perp}$ must be provided \cite{Molnar:2016vvu, BazowPhDThesis}. In addition, we must determine $\hat{\cal P}_\perp$. In kinetic theory
\begin{equation}
\hat{\cal P}_\perp\equiv-\frac{1}{2}\Delta^{\mu\nu}_{\perp}\hat{T}_{\mu\nu}
=\frac{1}{2}\left(\ed-\pL-\bar{\cal R}(\xi)(\ed-3\peq)\right)\;,\label{PTEQ}
\end{equation}
where the last term is the trace of the energy-momentum tensor. The $\bar{\cal R}$-function above depends only on the leading-order anisotropy of the system (measured by the macroscopic quantities $\ed/\pL$); we take the connection between $\bar{\cal R}$ and $\ed/\pL$ from kinetic theory. The relaxation equations can be obtained from the Boltzmann equation as~\cite{Molnar:2016vvu}
\begin{align}
d\Pit &= 
- \bar{\zeta}_l\, z_\mu D_l u^\mu - \bar{\zeta}_\perp\, \thetat-\frac{1}{\tau_\Pi}\left[\frac{1}{3}(1-\bar{\cal R})(\ed-3\peq)+\Pit\right]
-\IT_{\Pi}\;,
\label{dPi}
\\
%
d\pL &=  \bar{\zeta}^{l}_{l}\, z_\mu D_l u^\mu - \bar{\zeta}^{l}_{\perp}\, \tilde{\theta}
-\frac{1}{\tau_\pi}\left(\pL-\peq\right) - \IT_{L}\;,
\label{dPl}
\\
d\tilde{W}^{\mu}_{\perp z} &=
2 \bar{\eta}^W_{u} \, \Delta^\mu_{\perp,\nu} D_l u^\nu -2 \bar{\eta}^W_\perp \,z_\nu  \tilde{\nabla}^\mu u^\nu
-\frac{\WT^{\mu}_{\perp z}}{\tau_W}
-\IT^{\mu}_{W}
-G^{\mu}_{W}\;,
\label{eq:WTz_relEq}
\\
d\pit^{\mu\nu}_{\perp} &=
2\bar{\eta}\, \sigmat^{\mu \nu}-\frac{\pit^{\mu\nu}_\perp}{\tau_\pi}
-\IT^{\mu\nu}_{\pi}-G^{\mu\nu}_{\pi}.
\label{dpi}
\end{align}
We refer the interested reader to Ref.~\cite{BazowPhDThesis} for a complete listing and detailed discussion of all the terms on the right hand sides in Eqs.~(\ref{dPi})-(\ref{dpi}).
%
%
The ``transport coefficients" $\bar{\zeta}_\alpha$, $\bar{\kappa}_\alpha$, $\bar{\eta}_\alpha$ \big(where $\alpha$ is shorthand for all the sub-and superscripts attached to those letters in Eqs.~(\ref{dPi})-(\ref{dpi})\big) from kinetic theory are formally listed in Ref.~\cite{Molnar:2016vvu}. Following the procedure in \cite{Denicol:2014vaa} for standard viscous hydrodynamics, we evaluate them from kinetic theory for a system of light ($m/T{\,\ll\,}1$) particles in an ``anisotropic equilibrium'' state, keeping only the leading terms in $m/T$. After expressing $m/T$ in terms of $\ed$ (or $\peq$) and $c_s^2$ we take the latter from lattice QCD~\cite{BazowPhDThesis}.  

To numerically solve the conservation laws $\partial_{\mu}T^{\mu\nu}{\,=\,}0$ and relaxation equations (\ref{dPi})-(\ref{dpi}), we explicitly propagate $T^{\tau\mu}$, $\pL$, $\Pit$, and all components of $\WT^{\mu}_{\perp z}$ (4) and $\piT^{\mu\nu}$ (10) using a two-step Runge-Kutta scheme for the time integration and the Kurganov-Tadmor (KT) algorithm \cite{Kurganov} for the spatial derivatives. The other necessary ingredient in the algorithm is to reconstruct the inferred variables $\ed$ and $u^\mu$ from the numerically evolved quantities. This is more involved in anisotropic than in standard viscous hydrodynamics \cite{BazowPhDThesis}. We first define the known quantities $M^{\mu}\equiv T^{\tau\mu}-\piT^{\tau\mu}$. Then taking the combination $(u^3)^{2}M^{0}-u^{0}u^{3}M^3$ and using $u{\cdot}u=1$ and the orthogonality condition $z_{\mu}\WT^{\mu}_{\perp z}=0$, it can be shown that $z^\mu$ can be entirely written in terms of known quantities. Therefore, we can further define the known quantities $\bar{M}^{\mu}\equiv M^{\mu}-2\tilde{W}^{(\tau}_{\perp z}z^{\mu)}$, leading to a scalar equation for the magnitude of the transverse flow velocity:
\begin{equation}
u_{\perp}=\frac{\beta_\perp}{\beta_L+\pT(u_\perp)}\sqrt{1+u^2_\perp}.
\label{uTInferred_vah}
\end{equation}
It can be solved iteratively with the help of $\ed=\bar{M}^0-\tau\tilde{F}\bar{M}^{3}-u_\perp(1+u_\perp^2)^{-1/2}x\bar{M}_\perp$,
%
%
which enters in Eq.~(\ref{uTInferred_vah}) through ${\cal P}_\perp$ and the equation of state $\peq(u_\perp)\equiv\peq(\ed(u_\perp))$. $\beta_\perp$, $\beta_L$, and $\tilde{F}$ are known quantities \cite{BazowPhDThesis}. Knowing $u_\perp$, all of the components of the fluid velocity can be  reconstructed (with $x\equiv(1-\tilde{F}^2)^{1/2}$) from
\begin{equation} 
u^0 =\frac{1}{x}\sqrt{1+u_\perp^2}\, , \qquad
u^{1,2} = u_\perp\frac{\bar{M}^{1,2}}{\bar{M}^{}_\perp}\, , \qquad
u^3  =  \frac{\tilde{F}}{\tau x}\sqrt{1+u_\perp^2}.
\end{equation}

\vspace*{-5mm}
\section{Comparison}
\label{comp}
\begin{figure}[t!]
\begin{center}
    \begin{tabular}{ccc}
  \includegraphics[width=0.4\linewidth]{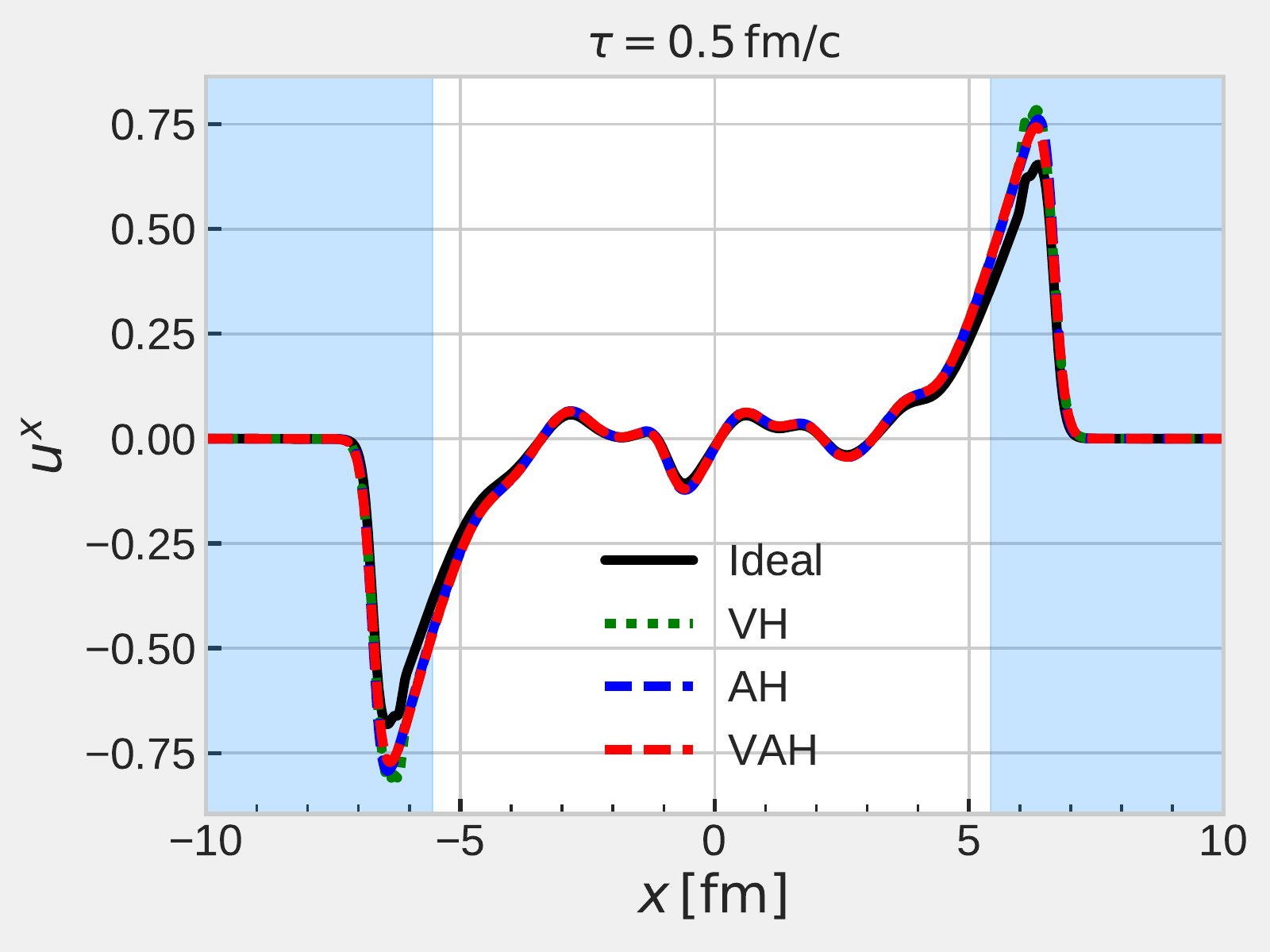} &
  \includegraphics[width=0.4\linewidth]{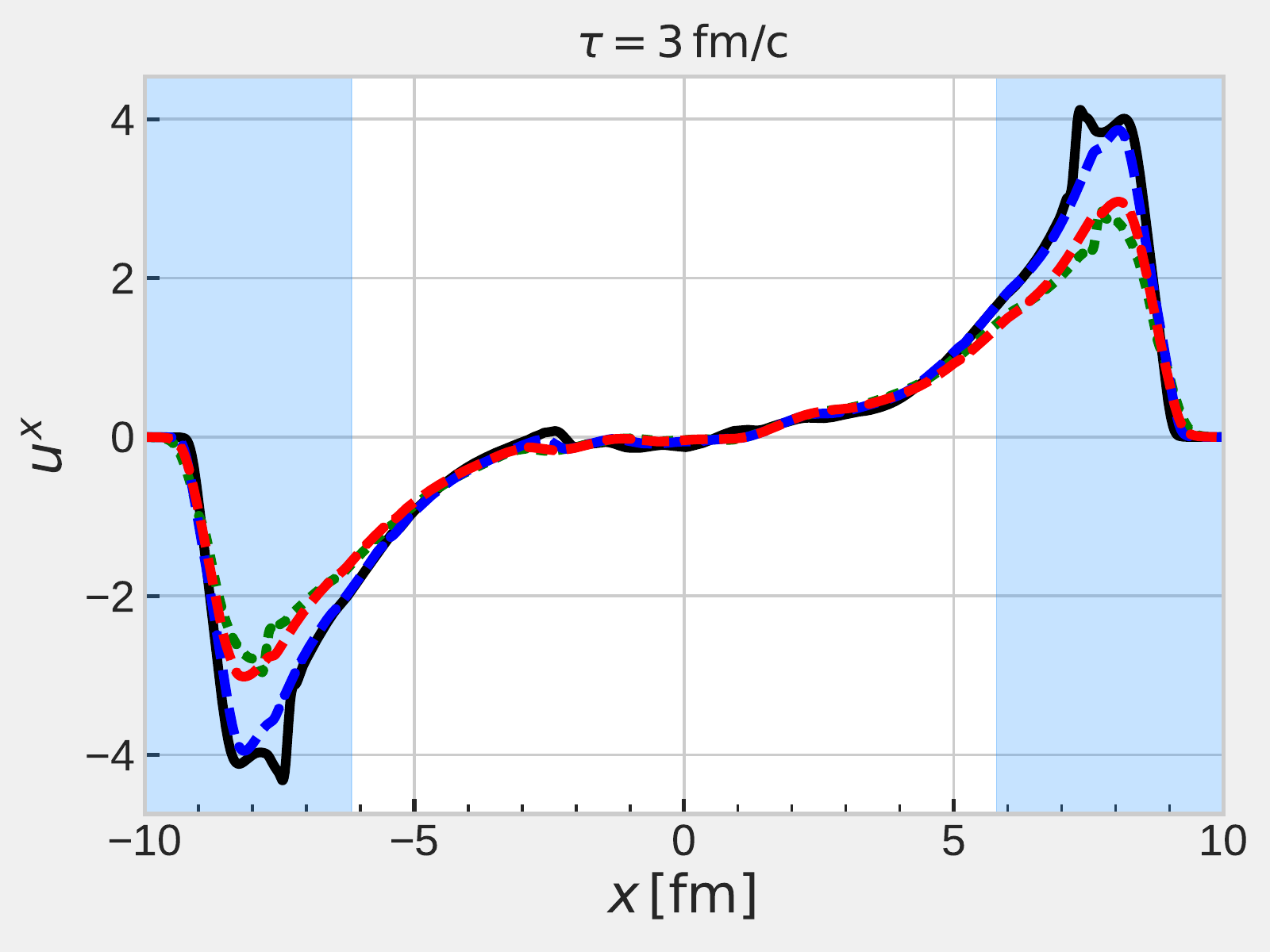} \\
    \includegraphics[width=0.4\linewidth]{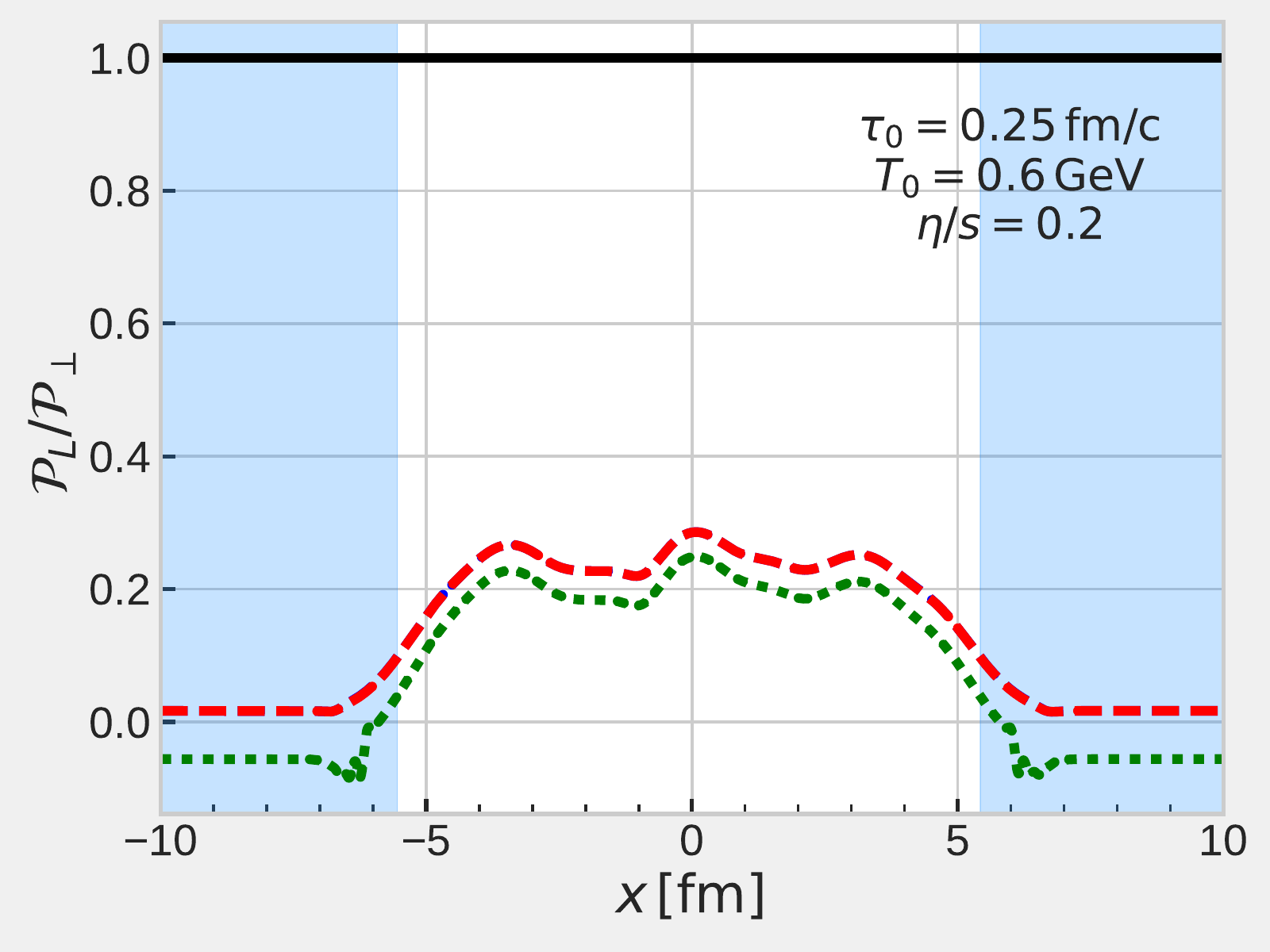} &
  \includegraphics[width=0.4\linewidth]{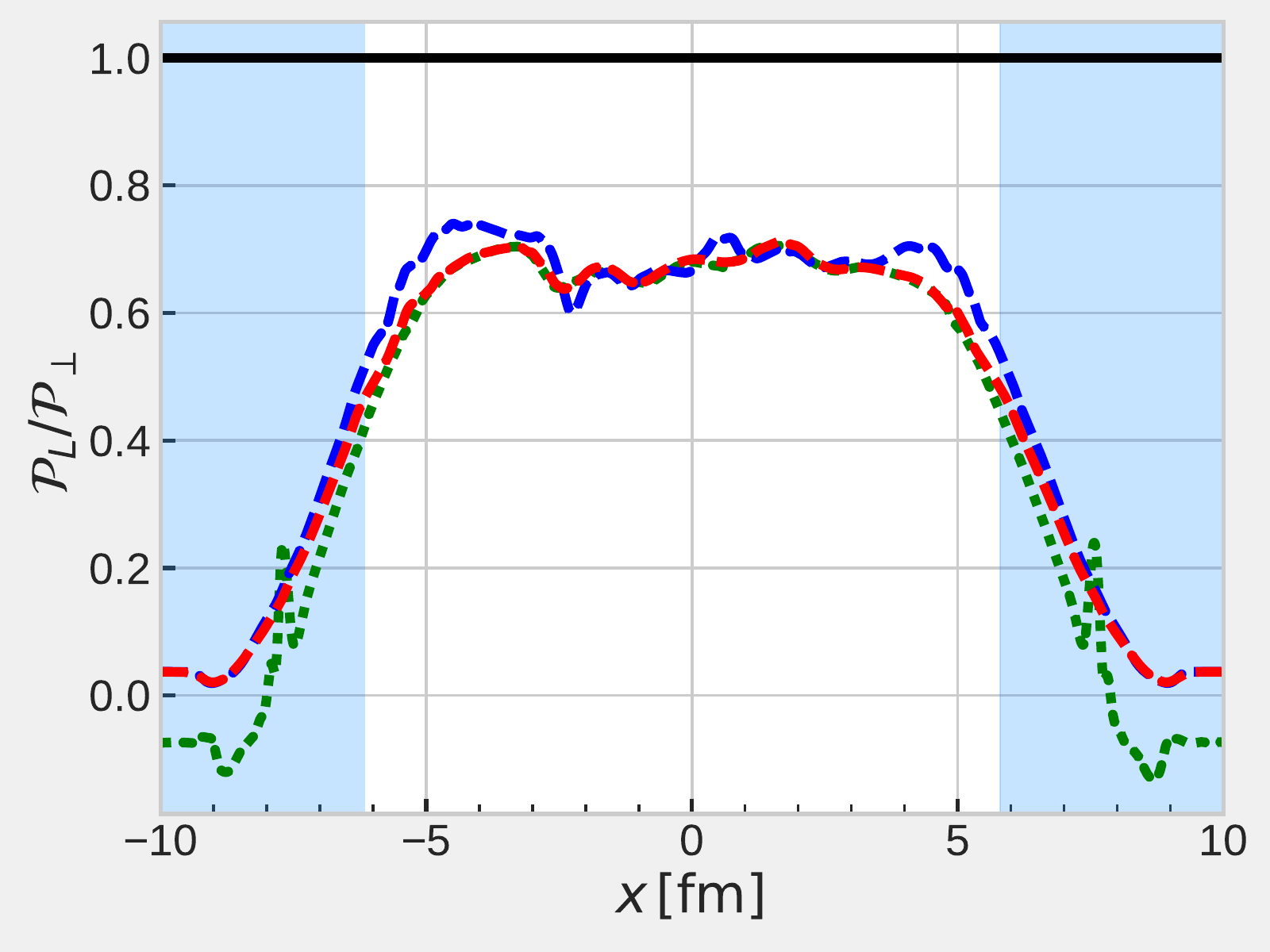}
    \end{tabular}
\hspace*{0.03\linewidth}
\caption{Comparison between ideal hydrodynamics (solid black line), viscous hydrodynamics (green line), anisotropic hydrodynamics (blue dotted line), and viscous anisotropic hydrodynamics (red line) plotted along the $x$ axis at a proper times $\tau{\,=\,}0.5$\,fm/$c$ (left panels) and  $\tau{\,=\,}3$\,fm/$c$ (right panels). The indivdual panels show the $x$ component of the fluid velocity and the pressure anisotropy ${\cal P}_L/{\cal P}_\perp$. The light shaded blue region represents points outside the isothermal freeze-out surface with $T_\mathrm{f}=155\,\mathrm{MeV}$. See text for details.}
\label{fig}
\vspace*{-7mm}
\end{center}
\end{figure}
As an example, we show in Fig.~\ref{fig} a comparison of the expansion velocity and macroscopic pressure anisotropy
along the $x$-axis between standard and anisotropic viscous hydrodynamics for a specific shear viscosity $\eta/s=0.2$. (Many more comparison plots, including the dynamics along the beam direction, can be found in \cite{BazowPhDThesis}.) The evolution is started at $\tau_0=0.25\,\mathrm{fm}/c$ with a spatially fluctuating (MC-Glauber) initial energy density profile normalized such that the average over many such profiles would correspond to an initial central temperature $T_0=0.6$\,GeV. The initial local momentum-space anisotropy is taken to be independent of position but relatively large, $\xi_0=10$. Fig.~\ref{fig} compares ideal hydrodynamics (solid black line) with viscous hydrodynamics (green dotted line, labeled VH), leading-order anisotropic hydrodynamics (blue dashed line, labeled VH), and second-order viscous anisotropic hydrodynamics (red dashed line, labeled VAH), both at a very early proper time value of $\tau{\,=\,}0.5$\,fm/$c$ (left panels) and at $\tau{\,=\,}3$\,fm/$c$ (right panels). At $\tau{\,=\,}0.5$\,fm/$c$ the system is dominated by longitudinal expansion, and the close agreement between AH and VAH reflects the smallness of the residual shear stress components $\WT^{\mu}_{\perp z}$ and $\piT^{\mu\nu}$. Comparing VH with VAH, we see about $15\%$ more pressure anisotropy ${\cal P}_L/{\cal P}_\perp$ in the center of the fireball and about $30\%$ larger anisotropies outside the freeze-out surface near $|x|{\,=\,}5$\,fm. In fact, VH gives negative ${\cal P}_L$ outside the freeze-out surface which is not the case for VAH. Later, at $\tau{\,=\,}3$\,fm/c, the transverse expansion has had enough time to grow to similar order as the longitudinal expansion rate, at least near the transverse edge of the fireball. At this time, the fluid velocities from VH and VAH agree well with each other, while AH is more similar to ideal hydrodynamics, indicating that the residual transverse shear viscous effects that have been ignored in AH are important for the transverse flow velocity profile. For the pressure anisotropy, the residual dissipative corrections in VAH, responsible for the difference between the dashed red (VAH) and blue (AH) lines, are much smaller than the dissipative correction in VH (responsible for the difference between VH and ideal hydrodynamics). This demonstrates the strong advantage of anisotropic over standard viscous hydrodynamic expansion schemes. 

%




\bibliographystyle{elsarticle-num}
\bibliography{Bazow_QM17}







\end{document}